\documentclass[11pt]{article}
\usepackage{moriond,epsfig}

\def\Journal#1#2#3#4{{#1} {\bf #2}, #3 (#4)}

\def\be{\begin{equation}}
\def\ee{\end{equation}}
\def\bea{\begin{eqnarray}}
\def\eea{\end{eqnarray}}

\begin{document}
\vspace*{4cm}
\title{The Large-Scale Structure view on the Galaxy-Quasar-AGN connection}

\author{Manuela Magliocchetti}

\address{INAF, Osservatorio Astronomico di Trieste, Via Tiepolo 11,\\
34100, Trieste, Italy}

\maketitle

\abstracts{Combined investigations of the clustering properties 
of galaxies of different spectral type and high-redshift quasars strongly suggest 
local ellipticals to be the parent population of optically bright Active 
Galactic Nuclei (AGN). However, the picture gets more blurred when one 
extends the analysis to that class of AGNs which show enhanced radio emission. 
Objects belonging to this class in fact are found 
to be associated with structures which are about an order of magnitude more massive 
than those that host radio-quiet AGNs. Also, masses for the black holes engines of 
radio-enhanced AGN emission turn out to be 
systematically higher than those which fuel 'normal' quasars. 
On the other hand, the level of radio-activity in radio-luminous objects does not seem 
to be connected with black hole/host galaxy mass, at variance with what found 
in the optical case. These results, together with evidences for different 
cosmological evolutions of different types of AGNs pose a serious challenge to 
all those models aiming at providing a unified picture for black hole-powered sources.}

\section{The Quasar-Galaxy Connection}
A careful analysis of the clustering properties of optically 
selected quasars has been performed by~\cite{porci}. These authors have 
concentrated on $\sim 15000$ objects taken from the 2dF Quasar Redshift 
Survey~\cite{croom} in the redshift range $0.8\le z\le 2.1$. 
Investigations of the projected spatial correlation function 
(see Figure 1) was performed by means of the so-called Halo Occupation Model. 
Its main quantity, the Halo Occupation Number (HON) $<N(M)>$, provides the 
mean 
number of sources brighter than some luminosity threshold hosted by a dark 
matter halo of chosen mass and, for halo masses greater than some 
chosen value $M_0$, can be written as follows: 
\begin{equation}
<N(M)>=N_0(M/M_0)^\alpha. 
\end{equation}
Combinations of values for the three parameters $M_0,N_0,\alpha$, 
presented in equation (1) 
have been considered so to match the observations in three different 
redshift ranges, in order to investigate any possible evolution of the 
quasar properties with look-back time. The main results, presented in 
Table~1 and in the left-hand panel of Figure~1, indicate that the typical 
minimum masses of haloes hosting optically selected quasars do not show great 
variations with redshift and lie in the narrow range 
$10^{12.2}-10^{12.7} M_\odot$. Also, the way the number of quasars 
hosted by a halo increases with its mass can be considered as constant 
($\alpha\simeq 0.7$) with good approximation. As a matter of fact, 
\cite{porci} found that the 2dF 
QSO clustering data can be easily described if one assumes values for the 
halo occupation number which stay constant in the whole redshift range 
covered by the data, denoting no evolution for at least some of the main 
structural properties connected to the presence and activity of quasars. 

\begin{table}[t]
\caption{Best-fitting values for the Halo Occupation Number (1) in the case of 
2dF early-type galaxies (column 1) and 2dF quasars at different redshifts 
(columns 2-4).}
\vspace{0.2cm}
\begin{center}
\begin{tabular}{|c|c|c|c|}
\hline
Early - $<z>\simeq 0.1$ & QSOs -  $<z>\simeq 1.1$ & QSOs - $<z>\simeq 1.5$ & 
QSOs - $<z>\simeq 1.9$ \\ \hline
$\alpha\simeq 1$& $\alpha=0.4^{+0.3}_{-0.3}$ &$\alpha=0.6^{+0.4}_{-0.4}$
& $\alpha= 1.1_{-0.7}^{+0.5}$\\
\hline
$\rm{log}[M_0/M_\odot]\simeq 12.6$& $\rm{log}[M_0/M_\odot]=12.2^{+0.2}_{-0.3}$ 
&$\rm{log}[M_0/M_\odot]\sim 12.5^{+0.2}_{-0.3}$& $\rm{log}[M_0/M_\odot]\sim 
12.7^{+0.3}_{-0.4}$\\ \hline
\end{tabular}
\end{center}
\end{table}

The HON estimates summarized in Table~1 are in extremely good agreement 
with those obtained by~\cite{maglio} in the investigation of the clustering 
properties 
of 2dF galaxies of early spectral type as measured by~\cite{madgwick}.
By again working within the HON framework, these authors 
indeed find that the projected correlation function of 
early-type galaxies (left-hand panel of Figure 2) can be very nicely 
described by assuming local elliptical galaxies to be hosted by haloes always 
more massive than $\sim 10^{12.6}M_{\odot}$, with a mean number increasing 
linearly with halo mass ($\alpha\sim 1$, cfr right-hand panel of Figure 2 and 
Table~1). We note that such an agreement between halo occupation properties 
of local ($z\le 0.2$) elliptical galaxies and intermediate-to-high redshift 
quasars 
is even more striking as the functional form adopted by~\cite{maglio} 
for $<N(M)>$ shows a number of differences when compared to that presented in 
equation (1).

\begin{figure*}
\vskip 6cm
\includegraphics{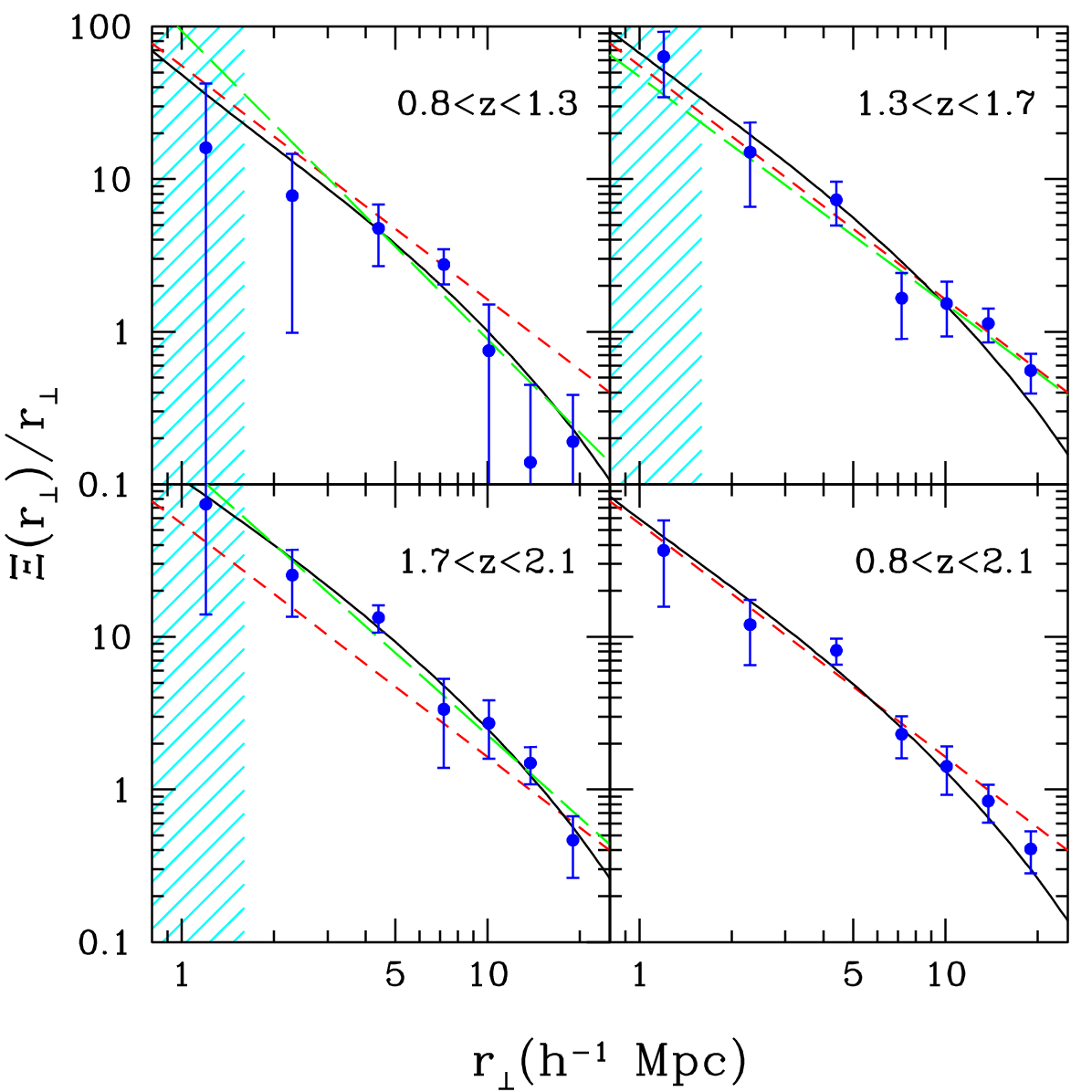}
\includegraphics{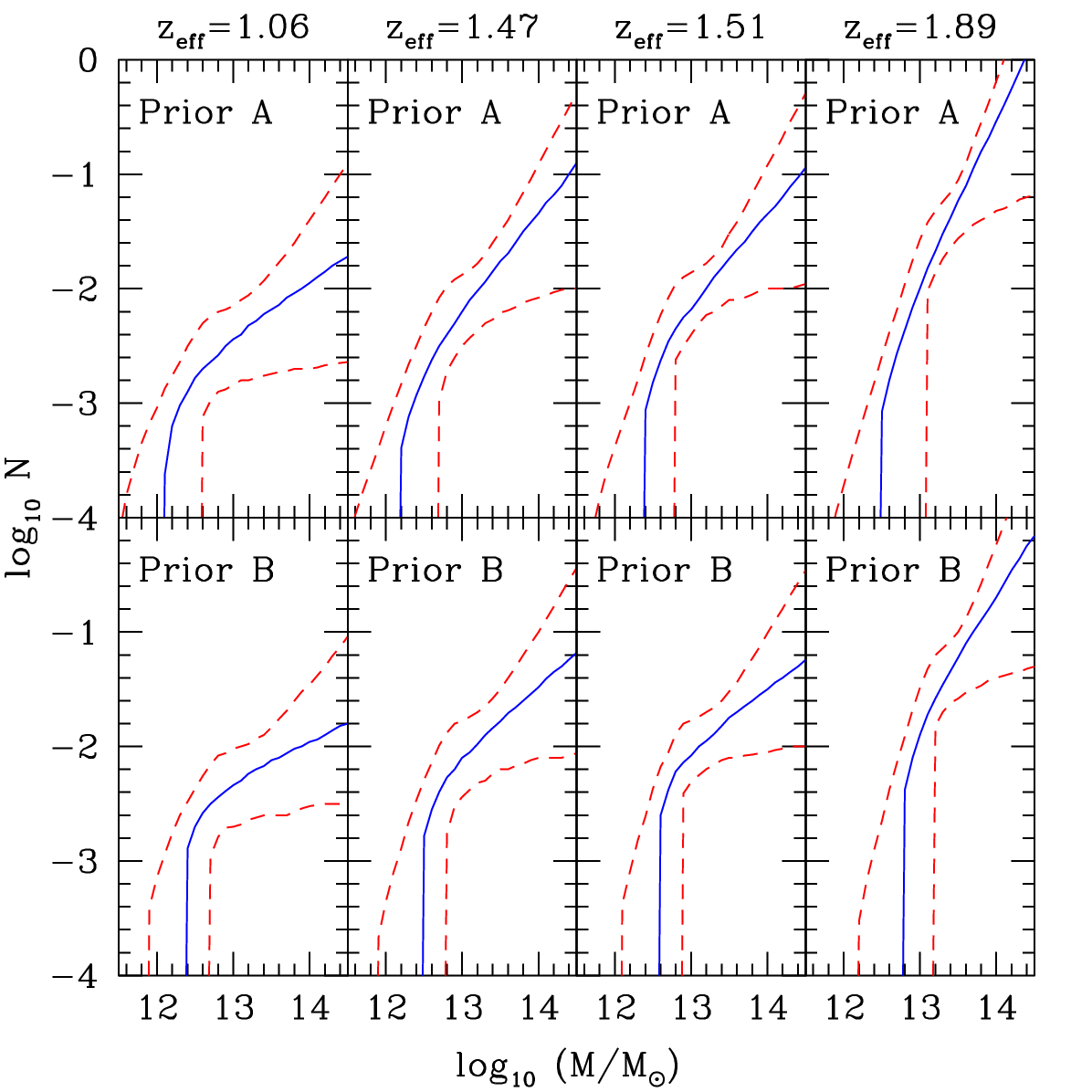}
\caption{Left-hand panel: Projected correlation function of 2dF 
quasars in different redshift bins as indicated in the panels. The dashed 
lines show the best power-law fit to the data, while the solid curves denote 
the best fits obtained within the HON framework. Right-hand panel: Best Halo 
Occupation Number (solid lines) and confidence intervals (dashed lines) 
as derived by$^1$.
\label{fig:QSO}}
\end{figure*}

\begin{figure*}
\vskip 5cm
\includegraphics{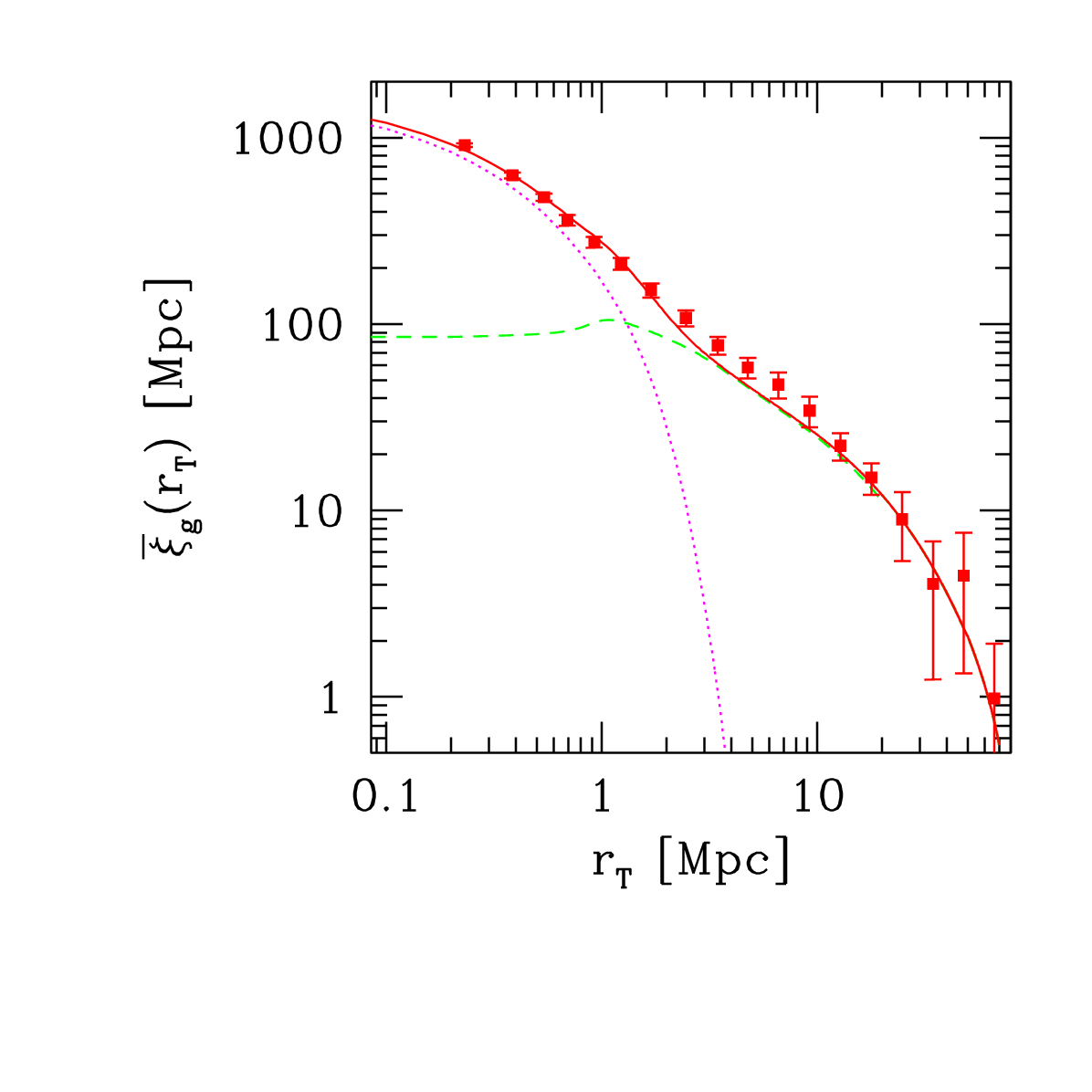}
\includegraphics{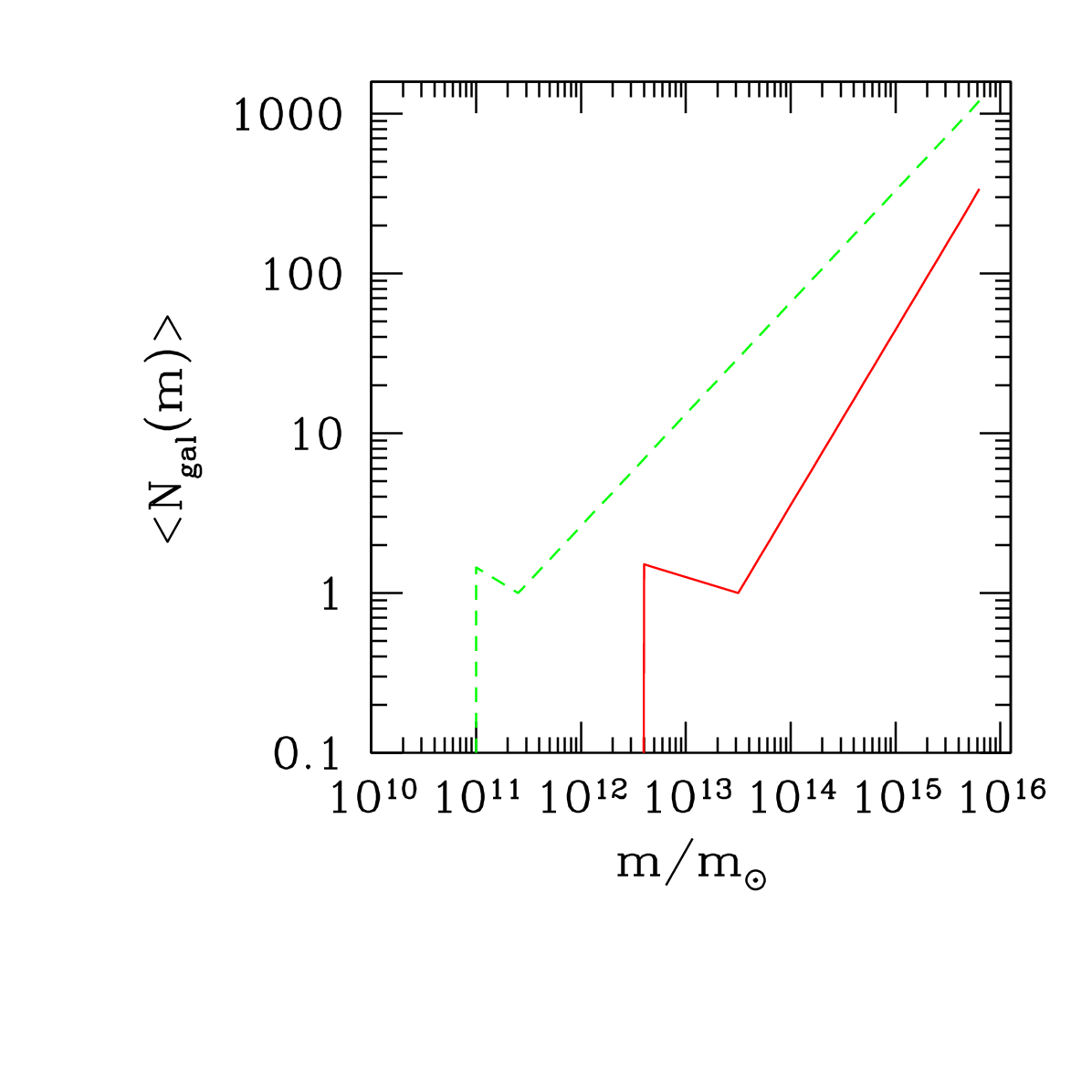}
\caption{Left-hand panel: Projected correlation function of local 2dF 
early-type galaxies. The solid line shows the best-fit to the data as 
calculated by$^3$. The corresponding 
best HON is indicated in the left-hand panel by the red-solid line.
\label{fig:early}}
\end{figure*}

The above results then imply that the structures which host QSOs at high 
redshifts are the same as those harbouring elliptical galaxies in the local 
universe. The picture that emerges purely from Large-Scale Structure 
studies is that of a strong evolutionary connection between local elliptical 
galaxies and high-redshift quasars, whereby the former class of sources can be 
identified as the parent population of bright, optically active AGNs. 
The chances to 
find an elliptical galaxy in its AGN phase is however low (in the redshift 
range considered by~\cite{porci} the ratio between the number of galaxies 
hosting a quasar to that of the whole galaxy population 
is $N_0\simeq 10^{-3}$), due to the limited life-time $t_Q$ 
of the quasar phenomenon. $t_Q$ is estimated by~\cite{porci} to be $\sim 7\cdot 10^7$~yr 
at $z\sim 2$, showing some hint for a decline at lower redshifts, possibly due 
to the more limited reservoir of gas needed to fuel the black hole (BH) 
hosted in the galaxy centres.

\section{Radio AGN}
Having dealt with optically selected quasars, we can now move on to 
investigating the properties of the broader class of Active Galactic Nuclei. 
In particular, we will concentrate our attention on radio-active sources, 
trying to tackle the following issues:

\begin{figure*}
\vskip 5cm
\includegraphics{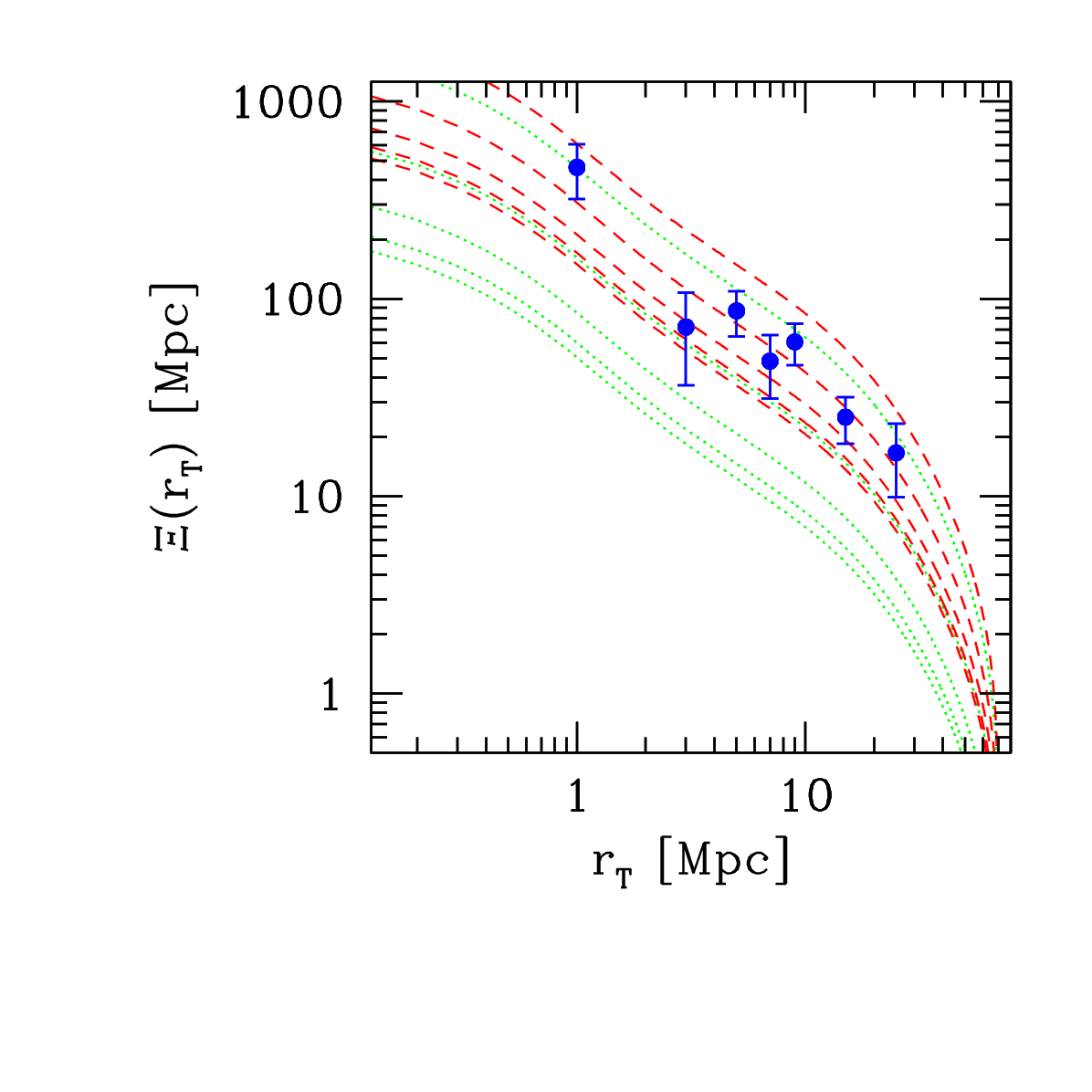}
\includegraphics{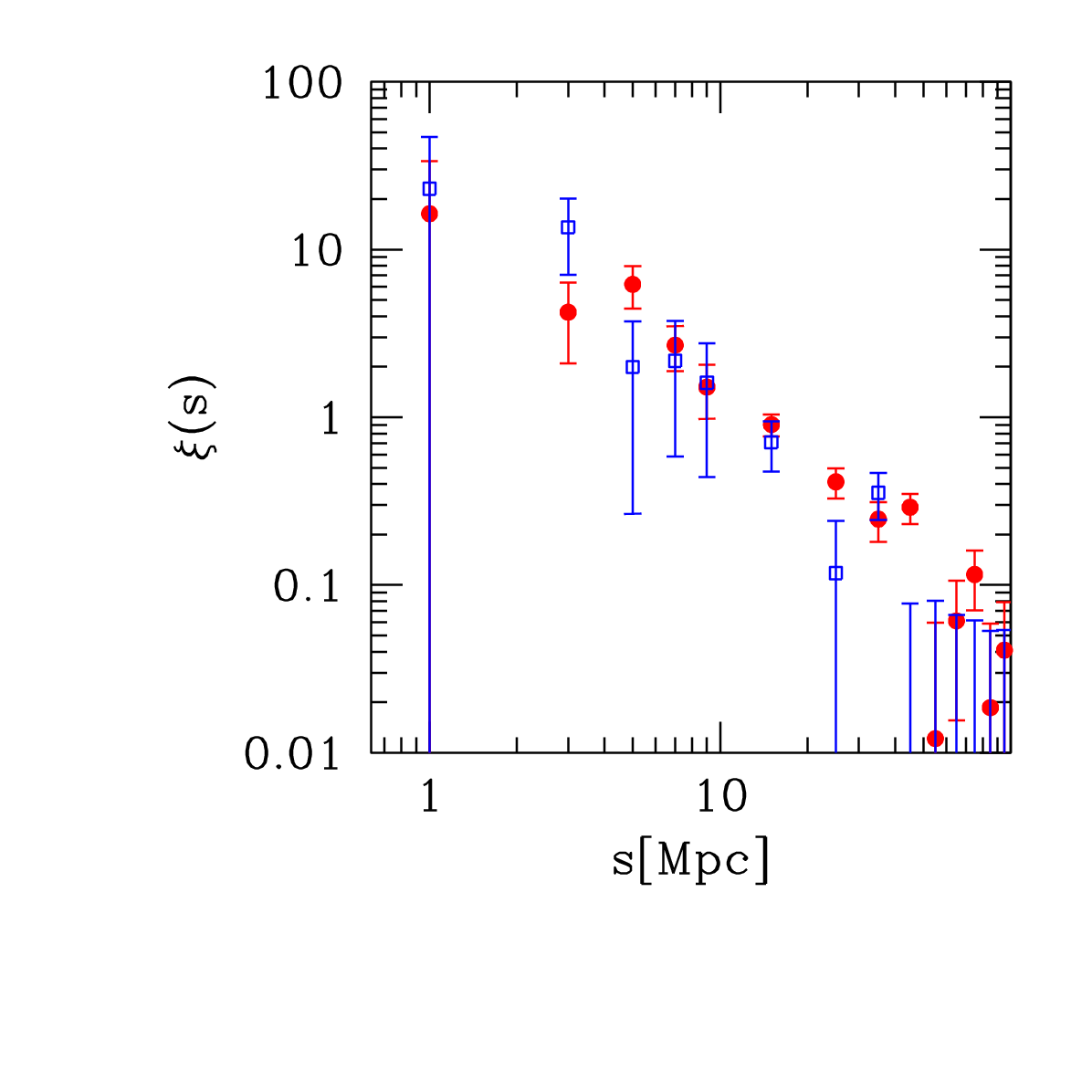}
\caption{
Left-hand panel: Projected correlation function of local 2dF radio galaxies 
as obtained by$^5$. The dashed lines indicate models for the correlation 
function of dark matter halos with masses respectively greater than 
$10^{10} M_\odot$ (bottom curve) to $10^{14} M_\odot$ (top curve).
Left-hand panel: Redshift-space correlation function of radio galaxies of 
different luminosity. Open squares are for for log(P)$>$22 [W/Hz/sr], 
filled dots for log(P)$<$22 [W/Hz/sr]
\label{fig:radio}}
\end{figure*}

\noindent
-- What is the connection between optical AGNs and radio sources?\\
-- Is the evolution of radio-active sources the same as that of optical AGNs?\\
-- Is radio-activity related to short-scale and/or large-scale processes?\\
-- And ultimately: what triggers radio-activity?

Once again, investigations of the large-scale properties of radio-active 
sources proves to be a very important diagnostic tool towards a correct 
interpretation of the AGN phenomenon. For instance, analyses of the 
clustering properties of the population of local radio galaxies performed 
by~\cite{maglio1} show that the correlation function of such sources 
has an amplitude which is about twice that found for 'normal' elliptical 
galaxies~\cite{madgwick}. Furthermore, comparisons with models (cfr left-hand 
panel of Figure~3) clearly indicate that local radio-active AGN have to 
reside in haloes more massive than $\sim 10^{13.5} M_\odot$. 
This figure is about a factor 10 higher than that found for optically selected 
quasars in Section 1. If we use the relation developped by~\cite{Ferrarese}, 
the above findings imply masses for the black holes powering the quasar 
phenomenon to be about an order of magnitude smaller than those fuelling 
radio galaxies ($\sim 10^8 M_\odot$ vs $\sim 10^9 M_\odot$).  

Another important piece of information can be obtained through the study of 
the clustering properties of sources with different radio luminosities. 
The results are illustrated in the right-hand panel of Figure~3 and 
show that, within the errors, differences in the redshift-space correlation 
function $\xi(s)$ of radio galaxies belonging to different classes of radio 
luminosity are negligible. If we then once again connect the clustering 
properties of bright and faint radio sources with the masses of the haloes 
which host them, and also refer to the results presented so far, we can 
conclude that there seems to be the need for {\it a threshold halo (BH) mass 
 to produce significant emission from AGN. However, once radio activity is 
onset, there is no evidence for a connection between radio luminosity and 
dark matter content (BH mass)}. We note that a strong back-up to the above 
statement comes from the results of~\cite{ben} who estimated BH 
masses directly from the spectra of a sub-sample of optically selected 
quasars with intense radio 
emission and found them not only to be independent on the level of radio 
activity, but also to be systematically more massive than those 
associated to the whole quasar population ($<M_{\rm BH}^{\rm RL}>\sim 10^{8.6} 
M_\odot$ vs $<M_{\rm BH}^{\rm RQ}>\sim 10^{8.3} M_\odot$). 

\begin{figure*}
\vskip 5cm
\includegraphics{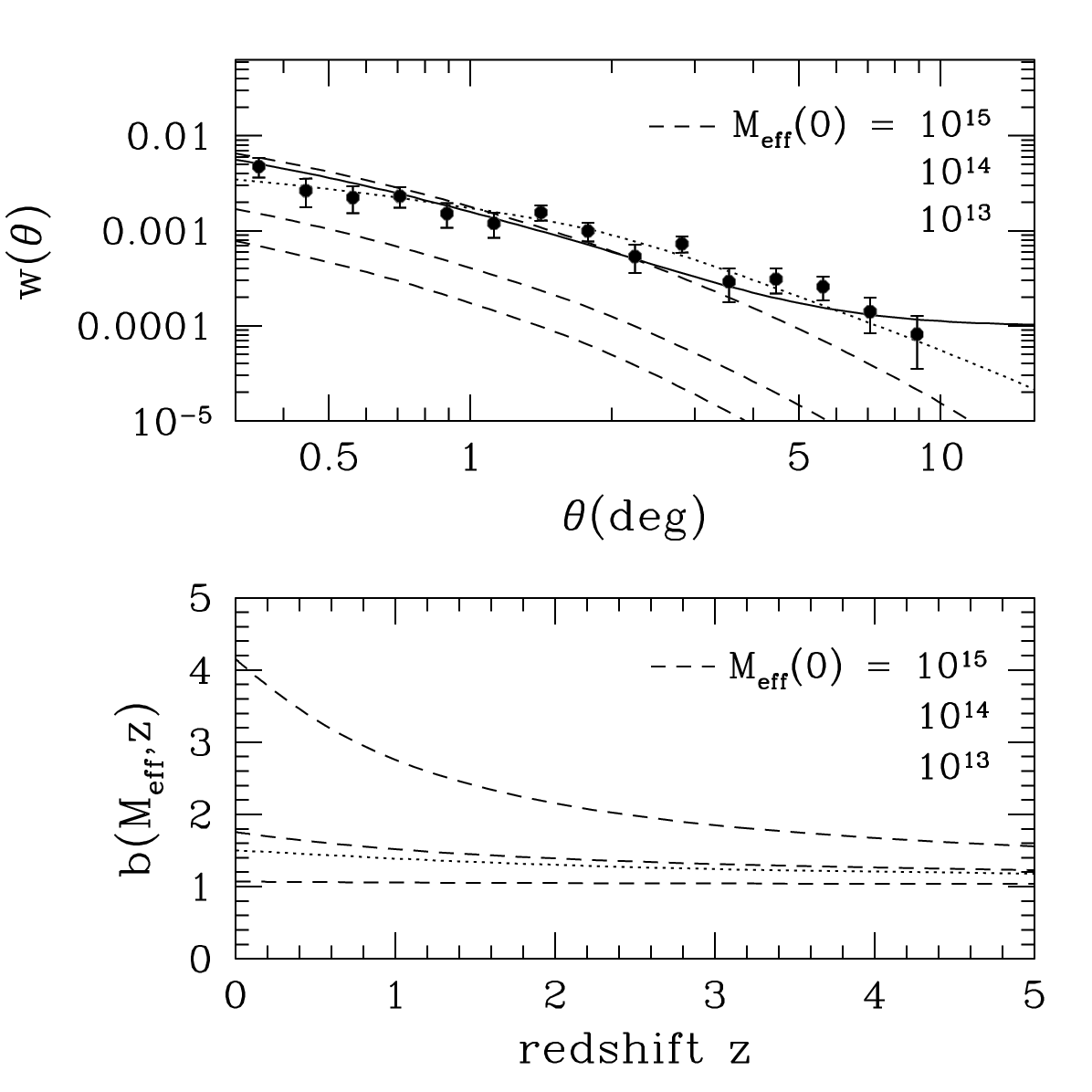}
\includegraphics{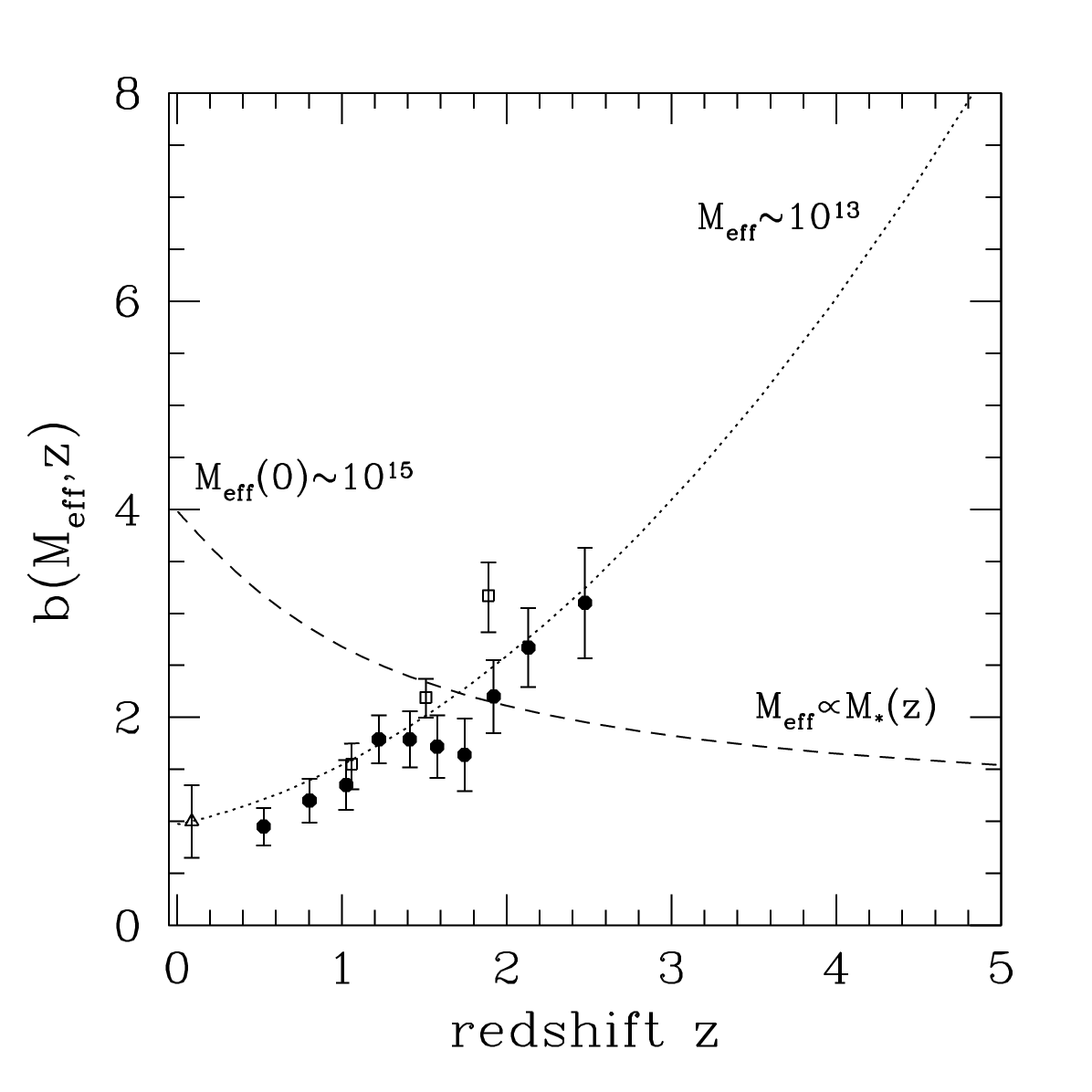}
\caption{
Left-hand panel: Angular two-point correlation function of NVSS 
radio sources as obtained by$^8$. The dashed lines indicate models for 
$w(\theta)$ based on halo masses which decrease with look-back time as 
$M_*$. From bottom to top these have been obtained for local 
values of the effective halo mass between $10^{13}$ and 10$^{15}$~
$h_0^{-1} M_\odot$. The bottom panel illustrates the corresponding 
time evolutions of the bias function. Left-hand panel: bias function 
$b(M_{\rm eff},z)$ as a function of redshift. Data points show the 
results obtained for optical quasars by$^{1,12,13}$, while the dotted curve 
illustrates the best-fit to the observations. The dashed curve reproduces 
the trend for the bias function found in the case of NVSS radio sources 
as reported in the bottom-right panel. Work from$^{10}$.
\label{fig:mattia}}
\end{figure*}

Some more interesting constraints on the behaviour of radio sources can be 
obtained by investigating the clustering properties of the whole radio 
population, regardless of their redshift distribution. The angular two-point 
correlation function $w(\theta)$ of sources brighter than 10~mJy as measured 
by~\cite{blake} in fact presents the puzzling 
behaviour of a power-law trend which extends up to angular scales of the order 
of $\sim 10$ degrees (left-hand panel of Figure 4).
The redshift distribution $N(z)$ of radio sources with 
S$_{1.4 \rm GHz}\ge 10$~mJy is predicted~\cite{DP} to be quite broad, spanning 
the $\sim 0-4$ redshift range, with a peak at $z \sim 1$. The maximum 
contribution to the angular correlation function $w(\theta)$ is then expected 
to originate from those sources which identify the $z \sim 1$ peak of $N(z)$. 
On the other hand, the spatial correlation function $\xi(r)$ for the dark 
matter is shown to become negative for comoving separations $r\ge 100$~Mpc 
(the exact value depending on the adopted cosmology) which, for a median 
redshift $z \sim 1$, corresponds to $\theta\sim 2$~degrees. 
It then follows that if one assumes constant masses for the dark matter halos 
which host radio sources as it was in the case of optically-selected quasars, 
it is impossible to explain the observed positive tail of $w(\theta)$ beyond 
$\sim 2$~degrees. We note that this problem is not alleviated either by adding 
to the total $w(\theta)$ a contribution from local star-forming galaxies or 
by adopting different halo masses. Even variations of the background cosmology 
(performed by considering 
different values for the parameters $h_0$, $\Omega_0$ and $\Omega_b$)
do not manage to reconcile models with observations (for a detailed approach 
to the above issues see~\cite{negrello}).

The only possible way out to this problem is to invoke a clustering which was 
weaker in the past. This can be done if we assume the characteristic mass 
of halos hosting radio sources to be proportional to $M_*$, i.e. the 
typical mass-scale when fluctuations collapse to form bond 
structures~\cite{Mo}. In this case, the predicted $w(\theta)$ provides a much 
better description to the data on all scales (cfr dashed curves in Figure 4). 
The best-fit correlation is provided in the standard WMAP cosmological scenario by 
a local mass for the halos hosting radio sources $M_{\rm eff}\sim 10^{15}
M_\odot$, showing once again that -- at least locally -- radio sources 
trace the distribution of the most massive structures in our universe such as 
clusters of galaxies.\\
Taken at face value, the above result points to different 
evolutionary properties for different classes of AGNs.
In fact, while optically selected quasars are associated to halos 
with effective masses $M_{\rm eff}\sim 10^{13} M_\odot$ which 
stay constant with look-back time up to the highest probed redshifts 
($z \sim 2.5$, cfr the right-hand panel in Figure 4), masses for those  
hosting radio sources are found to decrease with $z$. Indeed, radio sources seem 
to reside within the largest structures which collapse at any given epoch; 
in a narrow redshift interval around $z \sim 1.5$ these correspond to the hosts of 
optical quasars, while in the local universe they can be identified with 
groups and clusters of galaxies.

\section{Conclusions}
The results presented in this work and mainly based on Large-Scale Structure studies can be 
summarized as follows:
\begin{enumerate}
\item {There exists a strong evolutionary connection between high-z quasars and local 
elliptical galaxies.}
\item{ Local radio galaxies are associated to structures which are about a factor of 10 
   more massive than those which host optically-selected quasars. They are also most likely 
   fuelled by heavier BHs.}
\item{In radio-active sources there does not seem to be any connection between radio 
luminosity and BH/halo mass.}
\item{The clustering of radio sources is found to be weaker in the past. This implies
   radio-active objects to be hosted by halos with masses  
   which decrease with look-back time at variance with the quasar case which 
   requires $M_{\rm eff}\sim$~const at all redshifts probed by available surveys. 
   These two classes of objects 
   seem to be associated to the same kind of structures only at z~$\sim 1.5$.}
\end{enumerate}

The above conclusions raise a number of issues that need to be addressed 
whenever investigating possible connections -- both on large and on small scales --
between radio sources and optically-selected AGNs such as quasars.
In fact, these two classes of objects are found to reside in different structures, 
be fuelled by black holes of different masses and exhibit different cosmological 
evolutions, results which pose a serious challenge to
all those models aiming at providing a unified picture for black hole-powered sources.

\section*{Acknowledgments}
MM wishes to thank Cristiano Porciani, Mattia Negrello and Gianfranco De Zotti 
for their substantial contribution to the work presented in this conference.


\end{document}